# Role of the non-locality of the vector potential in the Aharonov-Bohm effect*

A. M. Stewart

Emeritus Faculty
The Australian National University
Canberra, ACT 0200, Australia.

**Abstract:** When the electromagnetic potentials are expressed in the Coulomb gauge in terms of the electric and magnetic fields rather than the sources responsible for these fields they have a simple form that is non-local *i.e.* the potentials depend on the fields at *every* point in space. It is this non-locality of classical electrodynamics that is primarily responsible for the puzzle associated with the static Aharonov-Bohm effect: that its interference pattern is affected by fields in a region of space that the electron beam never enters.

**1. Introduction**

One of the most fascinating and still controversial issues in physics is the Aharonov-Bohm effect [1,2], see also [3-10]. The experimental arrangement needed to observe the effect is shown schematically in Fig. 1. A long thin solenoid of radius $a$ is concentric with the $z$ axis which is directed out of the paper. The solenoid contains a magnetic flux $\Phi$. Inside the solenoid the magnetic field is $B^z = \Phi/\pi a^2$, outside the solenoid the field is $\mathbf{B} = 0$. From the left of the diagram an electron gun projects a beam of electrons towards a pair of slits, past the solenoid to impinge on a detector at the right of the diagram that detects the impact of each electron. The solenoid has an electrically conducting shield so that the electrons cannot penetrate into its interior and therefore cannot experience its magnetic field. The system is electrically shielded so that the electric field $\mathbf{E}$ within the system, other than that due to the presence of the electron beam, is zero also. When the solenoid is turned off the detector records a two-slit interference pattern resulting from the matter waves of the electron beam. When the solenoid carries a current the interference pattern shifts in the plane of the diagram with a flux period $2\pi\hbar/q$ where $\hbar$ is the Planck constant and $q$ the charge on the electron. The interference pattern develops its form even when the electrons are projected and detected one at a time.

The classical force $\mathbf{F}(\mathbf{r},t)$ that electric $\mathbf{E}$ and magnetic $\mathbf{B}$ fields exert on a charge $q$ at position $\mathbf{r}$ and with velocity $\mathbf{v}$ at time $t$ is given by the Lorentz expression [11]

$$\mathbf{F}(\mathbf{r},t) = q[\mathbf{E}(\mathbf{r},t) + \mathbf{v} \times \mathbf{B}(\mathbf{r},t)] \qquad . \qquad (1)$$

The issue that has caused puzzlement is why the interference pattern should be affected by the flux in the solenoid when the electron never experiences any classical force from a magnetic or electric field according to (1) because it never passes through any region of space in which the field $\mathbf{E}$ or $\mathbf{B}$ is non-zero. Although the electron beam induces screening charges on the inside surfaces of the interferometer, these charges do not experience the magnetic field of the solenoid either, because experiments with the solenoid surrounded by a superconducting shield, so that it is shielded from the magnetic fields produced by the electron beam, give the same result [12].

In Sec. 2 we examine the special features of the vector potential when it is expressed in the Coulomb gauge, particularly its non-local features. In Sec. 3 we apply these to the semi-classical interpretation [1,13,14] of the static Aharonov-Bohm effect that attributes it to a phase shift in the electron matter wave when it passes through a region of space in which the fields **E** and **B** are zero but in which the vector potential is non-zero. It is argued that the return path of the magnetic flux circuit has no influence on the effect. Section 4 gives a discussion of the results of the paper. In Appendix A it is shown that the potentials of the Coulomb gauge (9,10) are consistent with relations (2), in Appendix B it is shown that the potentials may be expressed in a single quasi-Lorentz invariant expression and in Appendix C there is a discussion of the surface integrals that are required to vanish at various points in the calculation. In Appendix D it is verified that equation (10) satisfies (6).

This paper discusses the standard interpretation of the Aharonov-Bohm effect - that it is caused by the electromagnetic potentials being non-zero in places where the fields themselves are zero [1,2]. Other interpretations of the effect have been suggested including those involving the formulation of quantum electrodynamics in terms of fields rather than potentials [7,8] (see also [9,10,15]) and many others, but discussion of these rival interpretations is beyond the scope of this paper.

**2 The Coulomb gauge**

It has not been found possible to construct a Lagrangian theory of the electromagnetic field, and therefore one that can be quantized canonically, without making use of the vector and scalar potential **A**(**r**,$t$) and $\phi$(**r**,$t$) defined through

$$\mathbf{B} = \nabla \times \mathbf{A} \qquad \mathbf{E} = -\nabla\phi - \frac{\partial \mathbf{A}}{\partial t} \tag{2}$$

where $\nabla$ is the gradient with respect to **r** ($\nabla'$ is the gradient with respect to **r**'). In classical electrodynamics these potentials are often used for calculational purposes [11,16]. The potentials are not unique: if the following transformation is made with the differentiable single-valued scalar field gauge function $\chi$(**r**,$t$)

$$\mathbf{A} \to \mathbf{A}' = \mathbf{A} + \nabla\chi \quad \text{and} \quad \phi \to \phi' = \phi - \partial\chi/\partial t \tag{3}$$

the electric and magnetic fields are unchanged. The gauge principle asserts that all physically measurable quantities are independent of the gauge function [17]. Following Panofsky [16] and Gubarev *et al.* [18] it is assumed that all scalar and vector fields, including the potentials, vanish at spatial infinity. Where the magnetic field is zero the vector potential may be expressed as the gradient of a scalar, because the curl of a gradient is zero: this is called a pure gauge field. Since the vector potential gives the same description of the magnetic field whatever arbitrary gradient is added to it, the question arises as to whether the vector potential possesses an irreducible kernel that encodes the full information about the magnetic field and which does not contain a part that can be expressed as a gradient. This question is answered affirmatively by considering the Helmholtz decomposition of a general 3-vector into its longitudinal and transverse parts [16,19-21]

$$\mathbf{A}(\mathbf{r},t) = -\nabla \int \frac{d^3r'}{4\pi} \frac{\nabla' \cdot \mathbf{A}(\mathbf{r}',t)}{|\mathbf{r}-\mathbf{r}'|} + \nabla \times \int \frac{d^3r'}{4\pi} \frac{\nabla' \times \mathbf{A}(\mathbf{r}',t)}{|\mathbf{r}-\mathbf{r}'|} \quad . \tag{4}$$

It is seen that a vector that has zero divergence has no component that can be expressed as a gradient and so satisfies this condition. The vector potential that obeys the gauge condition div**A** = 0 is said to be in the Coulomb gauge. Any arbitrary vector potential **A'** may be transformed into the Coulomb gauge by making a gauge transformation with the gauge function

$$\chi(\mathbf{r},t) = \int \frac{d^3r'}{4\pi} \frac{\nabla' \cdot \mathbf{A}'(\mathbf{r}',t)}{|\mathbf{r} - \mathbf{r}'|} \quad . \tag{5}$$

The Coulomb gauge has a minimal property [18,22]: it is the gauge that minimises the volume integral of $\mathbf{A}^2$ over all space. This result is obtained by squaring the first of (3) and using the relation $\nabla \cdot [\chi(\mathbf{r})\mathbf{A}(\mathbf{r})] = \chi(\mathbf{r})\nabla \cdot \mathbf{A}(\mathbf{r}) + \mathbf{A}(\mathbf{r}) \cdot \nabla \chi(\mathbf{r})$ (dropping the *t*) with a partial integration to eliminate the cross term and noting that the volume integral of the left hand side of the latter equation gives a surface term that vanishes. With the use of (4) [22] and the Coulomb gauge condition div**A** = 0 this gives

$$\int d^3r\, \mathbf{A}(\mathbf{r})^2 = \int d^3r \int d^3r' \frac{\mathbf{B}(\mathbf{r}) \cdot \mathbf{B}(\mathbf{r}')}{4\pi |\mathbf{r} - \mathbf{r}'|} + \int d^3r\, \nabla \chi(\mathbf{r}) \cdot \nabla \chi(\mathbf{r}) \tag{6}$$

where $\chi$ is the gauge function needed to change the potentials from the Coulomb gauge to another gauge: all terms are positive definite [22]. Also, the Coulomb gauge is a complete gauge as it has no residual gauge arbitrariness; if a gauge transformation is made that retains the gauge condition div**A** = 0 the gauge function has to satisfy $\nabla^2 \chi = 0$. The only solution to this equation that vanishes at infinity is $\chi = 0$. The irreducibility of the vector potential of the Coulomb gauge also follows. The curl of a vector field, such as the second term of (4), the transverse part, cannot be removed by a gauge transformation as it is impossible to express the curl of a vector field as the gradient of a scalar field $\chi$, as may be seen by taking the divergence of both which leads again to $\nabla^2 \chi = 0$. Accordingly, the only part of (4) that can be "gauged away", namely removed by making a gauge transformation, is the first term. Parts of the vector that have the form of a curl, such as the second term of (4) or (8) or (9), cannot be gauged away. They are irreducible.

The form of the scalar potential in the Coulomb gauge, expressed in terms of the sources that produce the field, has been known for a long time [11,16]

$$\phi(\mathbf{r},t) = \frac{1}{4\pi\varepsilon_0} \int d^3r' \frac{\rho(\mathbf{r}',t)}{|\mathbf{r} - \mathbf{r}'|} \tag{7}$$

where $\rho$ is the electric charge density and $\varepsilon_0$ the dielectric permittivity of space. The scalar potential is instantaneous in time. The vector potential of the Coulomb gauge in terms of the source currents has been obtained more recently by Jackson [23]

$$\mathbf{A}(\mathbf{r},t) = \frac{1}{4\pi\varepsilon_0} \nabla \times \int d^3r' \int_0^{R/c} d\tau\, \tau\, \mathbf{J}(\mathbf{r}',t-\tau) \times \mathbf{R}/R^3 \tag{8}$$

where $\mathbf{J}(\mathbf{r},t)$ is the electric current density, *c* is the speed of light and $R = |\mathbf{R}|$ with $\mathbf{R} = \mathbf{r} - \mathbf{r}'$. It has a more complicated form that is variably retarded in time, from zero retardation at $\tau = 0$ to full retardation at $\tau = R/c$.

However, when the potentials of the Coulomb gauge are expressed in terms of the *fields* rather than the sources they have the simple, symmetric, instantaneous and non-local forms

$$\phi(\mathbf{r},t) = \nabla \cdot \int d^3r' \frac{\mathbf{E}(\mathbf{r}',t)}{4\pi|\mathbf{r}-\mathbf{r}'|} \qquad (9)$$

$$\mathbf{A}(\mathbf{r},t) = \nabla \times \int d^3r' \frac{\mathbf{B}(\mathbf{r}',t)}{4\pi|\mathbf{r}-\mathbf{r}'|} \qquad . \qquad (10)$$

Equation (9) is obtained from (7) by using the Maxwell equation for div**E**, taking the divergence of the expression $\mathbf{E}(\mathbf{r}',t)/|\mathbf{r}-\mathbf{r}'|$ separately with respect to both $\nabla$ and $\nabla'$ and doing a partial integration with a vanishing surface term. Equation (10) is obtained from the Helmholtz decomposition (4) [24,25] and it may be verified that it satisfies the first of (2) by taking its curl [26]. In Appendix A it is confirmed that the potentials (9) and (10) are consistent with the defining equations (2). It is seen that the two potentials (9) and (10) encode their respective fields at every point in space at the same time.† Although the Coulomb gauge potentials show an apparent acausality it has been known for a long time [11] that they give rise to *fields* **E** and **B** that are manifestly causal. The non-local nature of the potentials was anticipated by Feynman [13]. The potentials (9) and (10) at time *t* are determinable (at some later time) to an arbitrary accuracy (within the limits allowed by quantum mechanics [27]) to the extent that the fields are measurable with detectors that have put in place in the past. It is only when the arbitrary gauge function $\chi$ of (3) is included in the potentials that they become undeterminable because of its presence.

The discussion in this section has been conducted with the assumption that the fields and potentials vanish at spatial infinity so that (4) is valid and all the terms of (6) remain finite. If the fields are made to be approximately uniform and constant over a volume of space, say that contains an experimental apparatus, then the potentials can be written as $\mathbf{A} = \mathbf{B} \times \mathbf{r}/2$ and $\phi = -\mathbf{E}\cdot\mathbf{r}$ in this limited region. These gauges [5,28] are useful for calculation of systems of limited extent such as those involving the Hall effect and Landau levels but cannot be physically realised in full as they diverge at infinity. Discussion of such divergent gauges lies outside the scope of this paper.

## 3. Consequences for the static Aharonov-Bohm effect

The vector potential for the static Aharonov-Bohm effect is usually obtained from the closed line integral around the solenoid given by Stokes' law [11]

$$\oint \mathbf{A}(\mathbf{r}) \cdot d\mathbf{r} = \Phi \qquad (11)$$

to be $\mathbf{A}(\mathbf{r}) = \hat{\boldsymbol{\theta}}\Phi/(2\pi\rho)$ where $\Phi$ is the flux enclosed by the path and with the assumption that the **A** field has cylindrical symmetry. This is a pure gauge field with $\mathbf{A}(\mathbf{r}) = \nabla g$ where $g = \Phi\theta/2\pi$ in the system of cylindrical coordinates $\mathbf{r} = \{\rho, \theta, z\}$. If $\theta$ is constrained to lie in the interval between 0 and $2\pi$, the multivalued scalar field *g* is written as $g = \Phi(n + \theta/2\pi)$ where *n* is the integral winding number.

However, the vector potential in the Coulomb gauge can also be obtained in a way that explicitly demonstrates its non-local nature [19,24]. We consider the flux path of an infinitely thin solenoid to be in the shape of a square of side 2*R*, where *R* >> *r*. One side of the square is directed along the *z* axis with one end at {0, 0, - *R*} and the other end at {0, 0, *R*}. We first calculate the contribution of this side to the vector potential near its centre at {0, 0, 0}. Using (10) with the field $\mathbf{B}(\mathbf{r}) = \hat{\mathbf{z}}\Phi\delta(x)\delta(y)$

$$\mathbf{A}(\mathbf{r}) = \frac{\Phi}{4\pi}\int dx'dy'dz'\frac{\delta(x')\delta(y')\hat{\mathbf{z}}\times(\mathbf{r}-\mathbf{r}')}{|\mathbf{r}-\mathbf{r}'|^3} \quad . \quad (12)$$

Because $\mathbf{r}' = \hat{\mathbf{z}}z'$ we get

$$\mathbf{A}(\mathbf{r}) = \frac{\Phi}{4\pi}\int_{-R}^{R}dz'\frac{(\hat{\mathbf{y}}x-\hat{\mathbf{x}}y)}{|\mathbf{r}-\mathbf{r}'|^3} \quad . \quad (13)$$

The numerator of the integrand is a vector in the $\theta$ direction of magnitude $\rho = (x^2 + y^2)^{1/2}$, the denominator is $[x^2 + y^2 + (z - z')^2]^{3/2}$. The integral then comes to

$$\mathbf{A}(\mathbf{r}) = \frac{\hat{\boldsymbol{\theta}}\Phi\rho}{4\pi}\int_{-R}^{R}\frac{dz'}{[\rho^2+(z-z')^2]^{3/2}} \quad . \quad (14)$$

The integral may be evaluated in closed form but it is more useful to express it as an expansion in *z*

$$\mathbf{A}(\mathbf{r}) = \hat{\boldsymbol{\theta}}\Phi\{\frac{1}{2\pi\rho[1+(\rho/R)^2]^{1/2}} - \frac{3}{4\pi R[1+(\rho/R)^2]^{5/2}}(\frac{\rho}{R})(\frac{z}{R})^2 + O[z^4]\} \quad . \quad (15)$$

When $R \gg \{\rho, z\}$ this attains the Stokes result. The same result can also be obtained with the multipolar gauge [11,29]. The contribution of the opposite side of the square is of order $\Phi/R$ and so is negligible. The contributions of the other two sides, which involve integrals from 0 to $2R$, also are of order $\Phi/R$. Contributions from the flux return path are therefore negligible and the standard assumption that the solenoid may be treated as infinitely long is justified. By giving the side of the square flux path the large but finite length $2R$ all the terms of equation (6) for the Coulomb gauge are found to remain finite. At large distances $\mathbf{A}$ will go as $1/r^2$ rather than $1/r$ and by giving the flux path a finite rather than a zero cross-section the right-hand side becomes finite.

    The interference pattern of the matter waves in the Aharonov-Bohm effect is a consequence of quantum mechanics [5,14]. The Schrödinger equation for a particle moving in a pure gauge potential is

$$\frac{1}{2m}(\mathbf{p}-q\nabla g)^2\psi_g = i\hbar\frac{\partial}{\partial t}\psi_g \quad (16)$$

where $\mathbf{p} = -i\hbar\nabla$. If the substitution

$$\psi_g = \psi_0 e^{iqg/\hbar} \quad (17)$$

is made then (16) becomes

$$\frac{1}{2m}\mathbf{p}^2\psi_0 = i\hbar\frac{\partial}{\partial t}\psi_0 \quad , \quad (18)$$

the equation of a free particle. It is seen that the effect of the pure gauge potential of the Aharonov-Bohm experiment is to add a phase to the particle's wavefunction. The path ABD through the top slit gains a phase of $-q\Phi/2\hbar$ (as the angle changes by $-\pi$), the path ACD through the bottom slit gains a phase of $q\Phi/2\hbar$. The phase difference between the paths, which gives rise to the interference pattern seen in the Aharonov-Bohm experiment, is $q\Phi/\hbar$. The addition of an arbitrary single valued gauge function $\chi$ contributes nothing to the phase shift because the line integral of such a field around a closed loop is zero. A multivalued gauge function, such as $g$, is not allowed as its addition would imply that magnetic flux, with observable physical consequences, had been added to the system.

**4. Discussion**

The vector potential of the electromagnetic field is found to contain an irreducible part that cannot be gauged away, the vector potential of the Coulomb gauge. This encodes sufficient information to describe the magnetic field, and no more than that. No residual gauge arbitrariness remains. The potentials of the Coulomb gauge may be expressed in simple forms in terms of their fields (9, 10). These forms are instantaneous in time and non-local in space; they depend only on the distribution of the electric and magnetic fields in space at the same time. In principle, the potentials are measurable if sufficient time is allowed for measurement signals to reach an observer at a later time. The arbitrariness of the gauge potentials resides with any gradient term that is added to the Coulomb gauge's vector potential; this can be eliminated by working on the Coulomb gauge.

The seeming non-locality of the Aharonov-Bohm effect, which has puzzled some, is attributed more to classical electrodynamics than to quantum mechanics because the only terms in the single-particle Schrödinger equation that are of a non-local nature are those involving the electromagnetic potentials. The general expression for the Schrödinger equation, of which (17) and (18) are special cases, is

$$\{\frac{1}{2m}[\mathbf{p}-q\mathbf{A}(\mathbf{r},t)]^2 + q\phi(\mathbf{r},t)\}\psi_s(\mathbf{r},t) = i\hbar\frac{\partial}{\partial t}\psi_s(\mathbf{r},t) \quad . \quad (19)$$

To describe the Aharonov-Bohm effect, (19) must be solved for $\psi_s$ everywhere in space, except within the solenoid where it is zero. At first glance (19) appears to be a purely local equation that depends only on position $\mathbf{r}$ and it is hard to see how it can "know about" the field in the solenoid to which it does not have access. However, when it is appreciated that the potentials $\mathbf{A}$ and $\phi$ encode information about the fields in other parts of space through (9) and (10), whether or not these parts of space are accessible to the wavefunction, it can be understood how fields in an inaccessible region can influence the wavefunction and hence the Aharonov-Bohm phase shift. The non-local nature of the electromagnetic potentials is essential for the Aharonov-Bohm effect to occur.

Feynman [13] has discussed the "reality" (his quotation marks) of the vector potential. He gives two different definitions of reality *(a) a real field is a mathematical function we use to avoid the idea of action at a distance.* His second definition is *(b) a real field is the set of numbers we specify in such a way that what happens at a point depends only on the numbers at that point.* The potentials of (9) and (10) appear to satisfy these definitions, but with the proviso that the set of numbers at one point (the potentials) implicitly encode the fields throughout all of space at the same time.† In this sense the potentials can be taken to be "real" fields. If the Lagrangian of the interaction of a charge with the electromagnetic field is written as

$$L = q[-\phi(\mathbf{r},t) + \mathbf{v}\cdot\mathbf{A}(\mathbf{r},t)] \tag{20}$$

it must be understood that this is a shorthand way of saying that the charge is interacting with the electromagnetic field not just at its position **r** but at every point in space at the same time through (9) and (10).

**Appendix A**

In this appendix it is verified that the potentials (9, 10) of the Coulomb gauge are consistent with relations (2). We differentiate (10) with respect to $t$ and use a Maxwell equation to get

$$\frac{\partial \mathbf{A}(\mathbf{r},t)}{\partial t} = -\int \frac{d^3r'}{4\pi} [\nabla'\times\mathbf{E}(\mathbf{r}',t)]\times\nabla'\frac{1}{|\mathbf{r}-\mathbf{r}'|} \quad . \tag{A1}$$

Next use the identity

$$\nabla'\times\frac{\nabla'\times\mathbf{E}(\mathbf{r}',t)}{|\mathbf{r}-\mathbf{r}'|} = \frac{\nabla'\times\nabla'\times\mathbf{E}(\mathbf{r}',t)}{|\mathbf{r}-\mathbf{r}'|} - [\nabla'\times\mathbf{E}(\mathbf{r}',t)]\times\nabla'\frac{1}{|\mathbf{r}-\mathbf{r}'|} \tag{A2}$$

and integrate by parts to get

$$\frac{\partial \mathbf{A}(\mathbf{r},t)}{\partial t} = -\int \frac{d^3r'}{4\pi} \frac{\nabla'\times\nabla'\times\mathbf{E}(\mathbf{r}',t)}{|\mathbf{r}-\mathbf{r}'|} \quad . \tag{A3}$$

By another vector identity this becomes

$$\frac{\partial \mathbf{A}(\mathbf{r},t)}{\partial t} = -\int \frac{d^3r'}{4\pi} \frac{\nabla'[\nabla'\cdot\mathbf{E}(\mathbf{r}',t)] - \nabla'^2\mathbf{E}(\mathbf{r}',t)}{|\mathbf{r}-\mathbf{r}'|} \quad . \tag{A4}$$

In the first term of (A4), using the identities

$$\nabla\frac{\nabla'\cdot\mathbf{E}(\mathbf{r}',t)}{|\mathbf{r}-\mathbf{r}'|} = \nabla'\cdot\mathbf{E}(\mathbf{r}',t)\nabla\frac{1}{|\mathbf{r}-\mathbf{r}'|} \tag{A5}$$

and, noting the change of sign arising from the interchange of $\nabla$ and $\nabla'$,

$$\nabla'\frac{\nabla'\cdot\mathbf{E}(\mathbf{r}',t)}{|\mathbf{r}-\mathbf{r}'|} = \frac{\nabla'[\nabla'\cdot\mathbf{E}(\mathbf{r}',t)]}{|\mathbf{r}-\mathbf{r}'|} - \nabla\frac{\nabla'\cdot\mathbf{E}(\mathbf{r}',t)}{|\mathbf{r}-\mathbf{r}'|} \tag{A6}$$

and an integration by parts, we get for the first term of (A4)

$$\frac{\partial \mathbf{A}(\mathbf{r},t)}{\partial t}|^1 = -\nabla\int \frac{d^3r'}{4\pi}\frac{\nabla'\cdot\mathbf{E}(\mathbf{r}',t)}{|\mathbf{r}-\mathbf{r}'|} = -\nabla\phi(\mathbf{r},t) \quad . \tag{A7}$$

To deal with the second term of (A4) we use the identity

$$\nabla' \cdot \frac{\nabla' E^i(\mathbf{r}',t)}{|\mathbf{r}-\mathbf{r}'|} = \frac{\nabla'^2 E^i(\mathbf{r}',t)}{|\mathbf{r}-\mathbf{r}'|} + [\nabla' E^i(\mathbf{r}',t)] \cdot \nabla' \frac{1}{|\mathbf{r}-\mathbf{r}'|} \qquad (A8)$$

to get, with a further integration by parts,

$$\frac{\partial A^i(\mathbf{r},t)}{\partial t}|^2 = -\int \frac{d^3 r'}{4\pi} [\nabla' E^i(\mathbf{r}',t)] \cdot \nabla' \frac{1}{|\mathbf{r}-\mathbf{r}'|} \qquad (A9)$$

With the use of another identity

$$\nabla' \cdot [E^i(\mathbf{r}',t) \nabla' \frac{1}{|\mathbf{r}-\mathbf{r}'|}] = [\nabla' E^i(\mathbf{r}',t)] \cdot \nabla' \frac{1}{|\mathbf{r}-\mathbf{r}'|} + E^i(\mathbf{r}',t) \nabla'^2 \frac{1}{|\mathbf{r}-\mathbf{r}'|} \qquad (A10)$$

we find with integration by parts

$$\frac{\partial A^i(\mathbf{r},t)}{\partial t}|^2 = \int \frac{d^3 r'}{4\pi} E^i(\mathbf{r}',t) \nabla'^2 \frac{1}{|\mathbf{r}-\mathbf{r}'|} \qquad (A11)$$

which, when using the relation $\nabla^2(1/|\mathbf{r}-\mathbf{r}'|) = -4\pi\delta(\mathbf{r}-\mathbf{r}')$, gives a result of $-E^i(\mathbf{r},t)$, so that the second term of (A4) is $-\mathbf{E}(\mathbf{r},t)$, thereby verifying the second of (2).

**Appendix B**

The potentials (9) and (10) may be expressed in the form of a single expression involving 4-vectors and tensors:

$$A^\mu(x) = \frac{\partial}{\partial x^\nu} \int \frac{d^4 x'}{4\pi} \frac{F^{\nu\mu}(x')}{|x-x'|} \delta(ct-ct') \qquad (B1)$$

where a 4-vector length is $x^\mu = \{ct, \mathbf{x}\}$, the potential 4-vector is $A^\mu = \{\phi/c, \mathbf{A}\}$ and the electromagnetic field tensor is

$$F^{\mu\nu} = \frac{1}{c}\begin{bmatrix} 0 & -E^1 & -E^2 & -E^3 \\ E^1 & 0 & -cB^3 & cB^2 \\ E^2 & cB^3 & 0 & -cB^1 \\ E^3 & -cB^2 & cB^1 & 0 \end{bmatrix} \qquad (B2)$$

giving $F^{i0} = E^i/c$ and $F^{ki} = \varepsilon^{kji} B^j$.

To obtain equations (9) and (10), first carry out the integration over $ct'$. This makes the times of $x$ and $x'$ equal. The potential $A^0$ in (10) comes from setting $\mu = 0$ in (B1)

$$A^0(x) = \frac{\partial}{\partial x^i} \int \frac{d^3 x'}{4\pi} \frac{F^{i0}(x')}{|\mathbf{x}-\mathbf{x}'|} \qquad (B3)$$

because the $\nu = 0$ term vanishes as $F$ is antisymmetric. Substituting for $F^{i0}$ we get (9). For $A^i$ we have

$$A^i(x) = \frac{\partial}{\partial x^k}\int \frac{d^3x'}{4\pi} \frac{F^{ki}(x')}{|\mathbf{x}-\mathbf{x}'|} \qquad (B4)$$

the differentiation by $x^0 = ct$ vanishing because the denominator of the integrand does not contain $t$. Substituting for $F^{ki}$ we have

$$A^i(x) = -\int \frac{d^3x'}{4\pi} \varepsilon^{ijk} B^j(x') \frac{\partial}{\partial x^k} \frac{1}{|\mathbf{x}-\mathbf{x}'|} \qquad (B5)$$

and, after applying a vector identity, this gives (10). Although (B1) is written in a form that is partly Lorentz invariant, it is not fully Lorentz invariant because of the delta-function term. If two space-time 4-vectors have equal times in one inertial frame they may not have equal times in a different frame.

**Appendix C**
At various points in the calculation it is required that volume integrals, that convert into surface integrals by means of one of Gauss' theorems, vanish. For example, in the derivation of (9), the volume integral over $d^3r$ of $\nabla'.[\mathbf{E}(\mathbf{r}',t)/|\mathbf{r}-\mathbf{r}'|]$ is transformed into a surface integral

$$\int d^3r' \nabla'.[\frac{\mathbf{E}(\mathbf{r}',t)}{|\mathbf{r}-\mathbf{r}'|}] = \oiint_{\infty} d\mathbf{S}'.[\frac{\mathbf{E}(\mathbf{r}',t)}{|\mathbf{r}-\mathbf{r}'|}] \qquad (C1)$$

at infinity with surface area element $d\mathbf{S}' = \hat{\mathbf{r}}' r'^2 d\Omega'$, where $d\mathbf{S}'$ is the vector surface area on a sphere of radius $r'$ directed in the direction of $\mathbf{r}'$ and $d\Omega$ is the solid angle. For the integral to vanish the integrand must approach zero faster than $1/r'^2$ as $r'$ goes to infinity. This is the case for static fields, but radiation fields go to zero more slowly and it must be considered if these satisfy the desired condition also.

The forms of the radiation fields of, for example, a charge oscillating in the $z$ direction ($\theta = 0$), are given by [30,31]. The dominant radiation terms at infinity, in spherical coordinates $\{r', \theta, \phi\}$ go as

$$E^{r'} \approx \cos\theta \sin(kr')/r'^2 \qquad E^\theta \approx \sin\theta \cos(kr')/r' \qquad E^\phi \approx 0 \qquad (C2)$$

$$B^{r'} \approx 0 \qquad B^\theta \approx 0 \qquad B^\phi \approx \sin\theta \cos(kr')/r' \qquad (C3)$$

where $k$ is the wave vector of the radiation. The vector cross product of $E^\theta$ and $B^\phi$ gives the Poynting vector, which is non-zero at infinity. However the surface integral arising from the left hand-side of (C1), because of the dot product, involves $E^{r'}$ which goes as $1/r'^2$, so the integral (C1) vanishes.

**Appendix D**
We show that (10) satisfies (6) with zero gauge function. Substituting (10) into (6) and dropping the $t$ parameter for convenience, we get for (6)

$$\int d^3r \, \mathbf{A}(\mathbf{r})^2 = \frac{1}{(4\pi)^2} \int d^3r' \int d^3r'' \int d^3r \, [\mathbf{B}(\mathbf{r}')\mathbf{x}\nabla \frac{1}{|\mathbf{r}-\mathbf{r}'|}] \cdot [\mathbf{B}(\mathbf{r}'')\mathbf{x}\nabla \frac{1}{|\mathbf{r}-\mathbf{r}''|}] \quad . \quad (D1)$$

Using a standard vector identity this becomes

$$\int d^3r \, \mathbf{A}(\mathbf{r})^2 = \frac{1}{(4\pi)^2} \int d^3r' \int d^3r'' \int d^3r \, \{\mathbf{B}(\mathbf{r}') \cdot \mathbf{B}(\mathbf{r}'')[\nabla \frac{1}{|\mathbf{r}-\mathbf{r}'|} \cdot \nabla \frac{1}{|\mathbf{r}-\mathbf{r}''|}] \\ -[\mathbf{B}(\mathbf{r}') \cdot \nabla \frac{1}{|\mathbf{r}-\mathbf{r}''|}][\mathbf{B}(\mathbf{r}'') \cdot \nabla \frac{1}{|\mathbf{r}-\mathbf{r}'|}]\} \quad . \quad (D2)$$

To evaluate the first term on the right-hand side of (D2) we use the identity

$$\nabla \cdot (\frac{1}{|\mathbf{r}-\mathbf{r}'|} \nabla \frac{1}{|\mathbf{r}-\mathbf{r}''|}) = \nabla \frac{1}{|\mathbf{r}-\mathbf{r}'|} \cdot \nabla \frac{1}{|\mathbf{r}-\mathbf{r}''|} + \frac{1}{|\mathbf{r}-\mathbf{r}'|} \nabla^2 \frac{1}{|\mathbf{r}-\mathbf{r}''|} \quad . \quad (D3)$$

Integrating the left-hand side of (D3) over $\mathbf{r}$ leads, by Gauss' law, to the surface integral $d\mathbf{S} = d\Omega \hat{\mathbf{r}} r^2 \cdot \hat{\mathbf{r}}/r^3$ which vanishes as $r \to \infty$. The second term on the right-hand side of (D3) provides a delta function $\delta(\mathbf{r} - \mathbf{r}'')$ which, when substituted into the first term on the right-hand side of (D2), gives (6) with zero gauge function. This is the required result.

Next we show that the second term on the right-hand side of (D2)

$$I_2 = -\frac{1}{(4\pi)^2} \int d^3r' \int d^3r'' \int d^3r \, [\nabla \cdot \frac{\mathbf{B}(\mathbf{r}')}{|\mathbf{r}-\mathbf{r}''|} \nabla \cdot \frac{\mathbf{B}(\mathbf{r}'')}{|\mathbf{r}-\mathbf{r}'|}] \quad (D4)$$

vanishes for radiation magnetic fields when integrated over $r \, r' \, r''$. We use the identity

$$\nabla \cdot [\frac{\mathbf{B}(\mathbf{r}')}{|\mathbf{r}-\mathbf{r}''|} \nabla \cdot \frac{\mathbf{B}(\mathbf{r}'')}{|\mathbf{r}-\mathbf{r}'|}] = \nabla \cdot [\frac{\mathbf{B}(\mathbf{r}')}{|\mathbf{r}-\mathbf{r}''|}] \nabla \cdot [\frac{\mathbf{B}(\mathbf{r}'')}{|\mathbf{r}-\mathbf{r}'|}] + \frac{\mathbf{B}(\mathbf{r}')}{|\mathbf{r}-\mathbf{r}''|} \cdot \{\nabla \nabla \cdot [\frac{\mathbf{B}(\mathbf{r}'')}{|\mathbf{r}-\mathbf{r}'|}]\} \quad . \quad (D5)$$

When integrated over $\mathbf{r}$, the left-hand side of (D5) becomes a surface integral over $\mathbf{r}$ which goes as $1/r$ and so vanishes at infinity. Using the relation

$$\nabla \nabla \cdot [\frac{\mathbf{B}(\mathbf{r}'')}{|\mathbf{r}-\mathbf{r}'|}] = \nabla[\mathbf{B}(\mathbf{r}'') \cdot \nabla \frac{1}{|\mathbf{r}-\mathbf{r}'|}] = [\mathbf{B}(\mathbf{r}'') \cdot \nabla] \nabla \frac{1}{|\mathbf{r}-\mathbf{r}'|} \quad , \quad (D6)$$

we get

$$I_2 = \frac{1}{(4\pi)^2} \int d^3r' \int d^3r'' \int d^3r \, \frac{\mathbf{B}(\mathbf{r}')}{|\mathbf{r}-\mathbf{r}''|} \cdot [\mathbf{B}(\mathbf{r}'') \cdot \nabla'] \nabla' \frac{1}{|\mathbf{r}-\mathbf{r}'|} \quad . \quad (D7)$$

Consider identity that describes the divergence of the product of a vector and a scalar

$$\nabla' \cdot \{\frac{\mathbf{B}(\mathbf{r}')}{|\mathbf{r}-\mathbf{r}''|} [\mathbf{B}(\mathbf{r}'') \cdot \nabla'] \frac{1}{|\mathbf{r}-\mathbf{r}'|}\} = \frac{\mathbf{B}(\mathbf{r}')}{|\mathbf{r}-\mathbf{r}''|} \cdot [\mathbf{B}(\mathbf{r}'') \cdot \nabla'] \nabla' \frac{1}{|\mathbf{r}-\mathbf{r}'|} \\ + [\nabla' \cdot \frac{\mathbf{B}(\mathbf{r}')}{|\mathbf{r}-\mathbf{r}''|}][\mathbf{B}(\mathbf{r}'') \cdot \nabla'] \frac{1}{|\mathbf{r}-\mathbf{r}'|} \quad . \quad (D8)$$

The second term on the right-hand side of (D8) vanishes because div'**B**' = 0. Accordingly, the integral $I_2$ vanishes provided that the integration over **r**' of the left-hand side of (D6) vanishes. The latter gives rise to a surface integral over **r**' at infinity. For the long-range field of electric dipole radiation (C3) the radial component of **B** vanishes, showing that integral $I_2$ is zero.

† Readers of a metaphysical turn of mind are advised not to jump too hastily to the conclusion that "the whole universe is instantaneously interconnected" or some such notion. The fields at different places at the same time are correlated because they are a result of sources acting at an earlier time. The influence of these sources is propagated by means of the Maxwell equations, which are causal.

Figure 1. Experimental set-up for the observation of the Aharonov-Bohm effect.

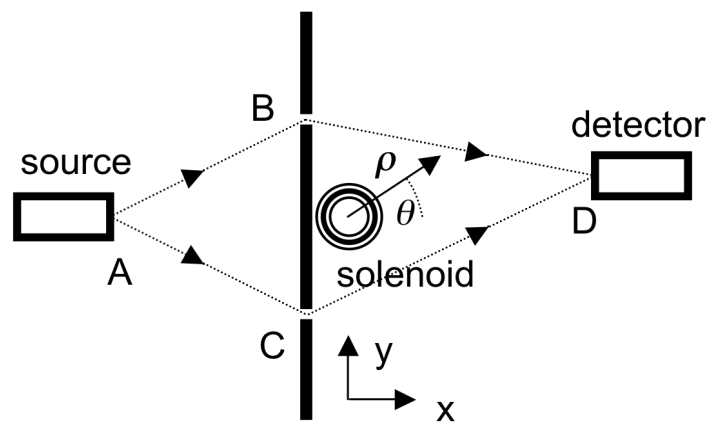